\documentclass{article}
\usepackage{hq2kpro}
\usepackage{graphicx}
\begin{document}
\title{
NEW RESULTS ON RARE AND FORBIDDEN SEMILEPTONIC $K^+$ DECAYS
}
\author{Peter Tru\"ol \\
{\em Physik-Institut, Universit\"at Z\"urich,
CH 8057 Z\"urich, Switzerland}}
\maketitle
\baselineskip=11.6pt
\begin{abstract}
Experiment E865\cite{E865} at the Brookhaven AGS was set up primarily to search
for the lepton flavour violating decay $K^+\rightarrow \pi^+\mu^+e^-$ 
($K_{\pi\mu e}$) with high sensitivity.
The flexibility of the apparatus allowed also to obtain more than an order
of magnitude larger than previously available event samples on the following
decay modes:
$ \pi^+e^+e^-\;(K_{\pi ee})$, $\pi^+\mu^+\mu^-\;(K_{\pi\mu\mu})$,
$\pi^+\pi^-e^+\nu_e\;(K_{e4})$, $\mu^+e^+e^-\nu_\mu$, and $e^+e^+e^-\nu_e$.
The report focusses on the $K_{\pi\mu e}$ results and those
on other lepton flavour violating decays as well as the
$K_{e4}$ data, from which a new, quite precise
value for the $s$-wave $\pi\pi$ scattering length can be deduced.
\end{abstract}
\baselineskip=14pt
\section{The E865 experiment: goals and techniques}
\label{sec:intro}
The search for a violation of the apparent, but theoretically not well founded
lepton number conservation law continues at several laboratories around the
world. Improving the sensitivity beyond existing limits for the
decays $\mu^+\rightarrow e^+\gamma$ ($<1.2\times 10^{-11}$)\cite{Brooks99}
and $\mu^-A\rightarrow e^-A^\prime$ ($<1.7\times 10^{-12}$)\cite{Kaulard98} 
is the goal of ongoing or planned experiments at the Paul Scherrer Institut
(PSI)
and Brookhaven National Laboratory (BNL)\cite{LFV00}. The observed evidence
for neutrino oscillations\cite{Kamiokande00} has further stimulated the
interest in this area. Since extensions of the
Standard Model like supersymmetry, leptoquarks, technicolor 
and other horizontal
gauge boson models have been proposed
which allow the possibility of flavour changing neutral 
currents and lepton family violating interactions, the forbidden $K$ decays 
$K_L^0\rightarrow\mu^\pm e^\mp$ ($<4.7\times 10^{-12}$)\cite{Ambrose98}  
and $K^+\rightarrow \pi^+\mu^+e^-$ 
($K_{\pi\mu e} <2.0\times 10^{-10}$)\cite{Lee90} 
were also investigated. E865 was set up to search 
for the latter decay with improved sensitivity.
\begin{figure}[h]
\includegraphics[width=53mm,angle=90]{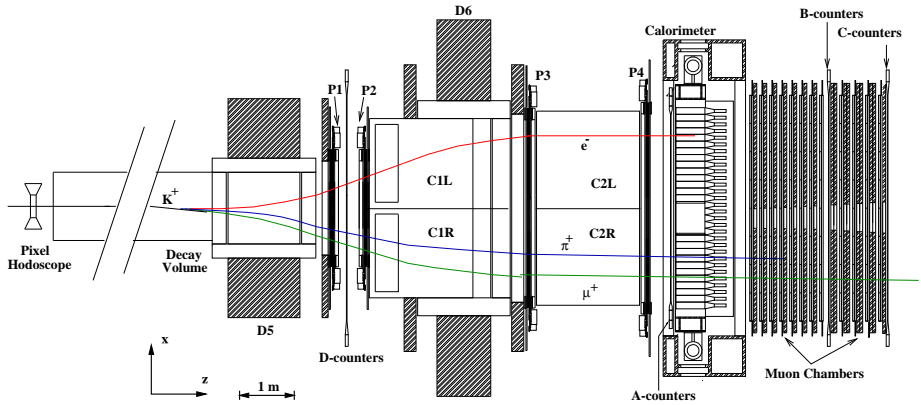}
\caption{\it Plan view of the E865 spectrometer.\label{fig:det}}
\end{figure}

The E865 detector (see Fig.~\ref{fig:det}) resided in an 
unseparated 6 GeV/$c$ beam of $1.5\times 10^7$ $K^+$ accompanied by
about $3 \times 10^8$ $\pi^+$ and $p$ per AGS cycle. About 8\% of the
$K^+$ decayed in the 5 m long decay volume. The decay products were
separated by charge by a first dipole magnet and then momentum analysed
by a second dipole magnet sandwiched between two pairs of
proportional chambers. The wire chambers with four planes each were
kept at lower voltage in the region were the beam passes. This arrangement
yields a resolution of $\sigma_p/p^2\approx 0.003$, where
the momentum $p$ of the decay products ($\pi^\pm,\mu^\pm$ or $e^\pm$)
ranges between 0.6 and 3.5 GeV/$c$. Four gas filled \v{C}erenkov
counters were positioned inside and after the spectrometer
magnet for $e^\pm$ identification. An electromagnetic calorimeter of
Shashlyk type\cite{Atoyan92} further helped to separate $e^\pm$ from other
decay products. It consists of 600 modules, each 15 radiation lengths 
deep in beam direction. It is followed by an array of 12 planes of 
proportional tubes separated by iron plates used to discriminate pions 
against muons.

The event trigger is based on the information from four 
scintillator hodoscopes, D and A before and after the spectrometer, B and 
C in front and in the middle of the muon stack. In the lowest trigger level
three charged particles are required, each identified by a cluster in the 
calorimeter and a signal from the A-counter in front of it. Depending on 
the final state the particle identification signals and topological 
information is added at higher levels, e.g. a $K_{\pi\mu e}$ candidate 
is given by an $e^-$ (\v{C}erenkov on beam left filled with H$_2$
at atmospheric pressure) and 
two particles on the beam right, one traversing the muon stack, vetoed 
with signals from the CH$_4$ filled \v{C}erenkov on beam right. 
The flexibility of the trigger and particle identification system
allowed to accumulate high statistics data samples either
concurrently with $K_{\pi\mu e}$ high intensity running or in
dedicated runs at lower intensity on the decays 
$\pi^+e^+e^-\;(K_{\pi ee})$, $\pi^+\mu^+\mu^-\;(K_{\pi\mu\mu})$,
$\pi^+\pi^-e^+\nu_e\;(K_{e4})$, $\mu^+e^+e^-\nu_\mu$, $e^+e^+e^-\nu_e$
for physics analysis and on 
$\pi^+\pi^0\;(e^+e^-\gamma)\;(K_{Dalitz})$ and $\pi^+\pi^+\pi^-\;(K_\tau)$
for normalisation.

The report focusses on the $K_{\pi\mu e}$ results and those
on other lepton flavour violating decays as well as the
$K_{e4}$ data, from which a new, quite precise
value for the $s$-wave $\pi\pi$ scattering length can be deduced. This
quantity plays a central role in the low energy effective theory of strong
interactions - chiral perturbation theory (ChPT)\cite{Gasser83,Colangelo00}. 
Both the flavour changing
neutral current decays $K_{\pi\mu\mu}$ and $K_{\pi ee}$, which
will not be discussed here in detail, are known to be dominated in the
Standard Model by long-distance effects involving
one-photon exchange, and are of interest for ChPT tests, too. Our 
published results\cite{Appel99,Ma00} for these decays  include precise
branching ratios and form factor information, establish firmly the vector 
nature of the interaction, resolve the discrepancy of older data with   
$e\mu$ universality, and provide evidence for the existence of a pion 
loop term calculated in next-to-leading order ChPT.

\section{Lepton flavour number violating decays}
\label{sec:lfv}
 
\begin{table}[htb]
\caption{\it Upper limits for lepton flavour number violating
decays established by experiment E865\cite{Appel00a,Appel00b}:
LFNV: violation of lepton flavour number conservation ($L_\mu$, $L_e$);
LNV: violation of total lepton number conservation ($L=L_\mu+L_e$);
$G_i$: number of quarks and leptons of generation $i$\cite{Cahn80};
GV: violation of generation number conservation. (PDG:
particle data group\cite{Caso00})}
\centering
\vskip 0.1 in
\begin{tabular}{|l|l|l|l|}\hline
$K^+$ decay & Branching  & Data & Remarks \\
mode & ratio &  & LFNV \\\hline
{$\pi^+\mu^+e^-$} & {$<3.9\cdot 10^{-11}$} &  
$K_{\pi\mu e}$ (1996) & 
$\Delta L=0$ \\
$K_{\pi\mu e}$   &  {$ <2.1\cdot 10^{-10}$} & $K_{\pi\mu e}$ (1995) & 
$\Delta L_e=-\Delta L_\mu=1$ \\
   &  $<2.0\cdot 10^{-10}$ & [Lee 1990] & 
$\Delta G_1=\Delta G_2=0$ \\\cline{2-4}
   &  {$<2.8\cdot 10^{-11}$} & \multicolumn{2}{|l|}{Combined} \\
\cline{2-4}
   &  $<1.0\cdot 10^{-11}$ & \multicolumn{2}{|l|}{Expected (1998)} 
\\\hline\hline
{$\pi^+e^+\mu^-$} & $<7.0\cdot 10^{-9}$ & PDG & GV, $\Delta L=0$ \\
$K_{\pi e\mu}$ & {$<5.1\cdot 10^{-10}$} &  $K_{e4}$ (1997) 
& $\Delta L_\mu=-\Delta L_e=1$
\\
&&&$\Delta G_2=-\Delta G_1=2$\\\hline\hline
{$\pi^-\mu^+e^+$} & $<7.0\cdot 10^{-9}$ & PDG & 
LNV, GV, $\Delta L=-2$ \\
$K_{\mu e\pi}$&{$<4.9\cdot 10^{-10}$}& $K_{e4}$ (1997) &
$\Delta L_\mu=\Delta L_e=-1$ \\
&&&$\Delta G_2=0,\; -\Delta G_1=2$\\\hline
{$\pi^-e^+e^+$} & $<1.0\cdot 10^{-8}$ & PDG & LNV, GV \\
$K_{ee\pi}$&{$<6.3\cdot 10^{-10}$}&  $K_{e4}$ (1997) & 
$\Delta L=\Delta L_e=-2$\\
&&&$\Delta G_2=1,\;-\Delta G_1=3$\\\hline
{$\pi^-\mu^+\mu^+$} & $<1.5\cdot 10^{-4}$ &PDG & LNV, GV \\
$K_{\mu\mu\pi}$&{$<3.0\cdot 10^{-9}$} & $K_{\pi\mu\mu}$ (1997) 
&$\Delta L=\Delta L_\mu=-2$\\
&&&$\Delta G_2=\Delta G_1=-1$\\\hline
\end{tabular}
\label{tab:lfv}
\end{table}

\begin{figure}[htbp]
\includegraphics[width=66mm]{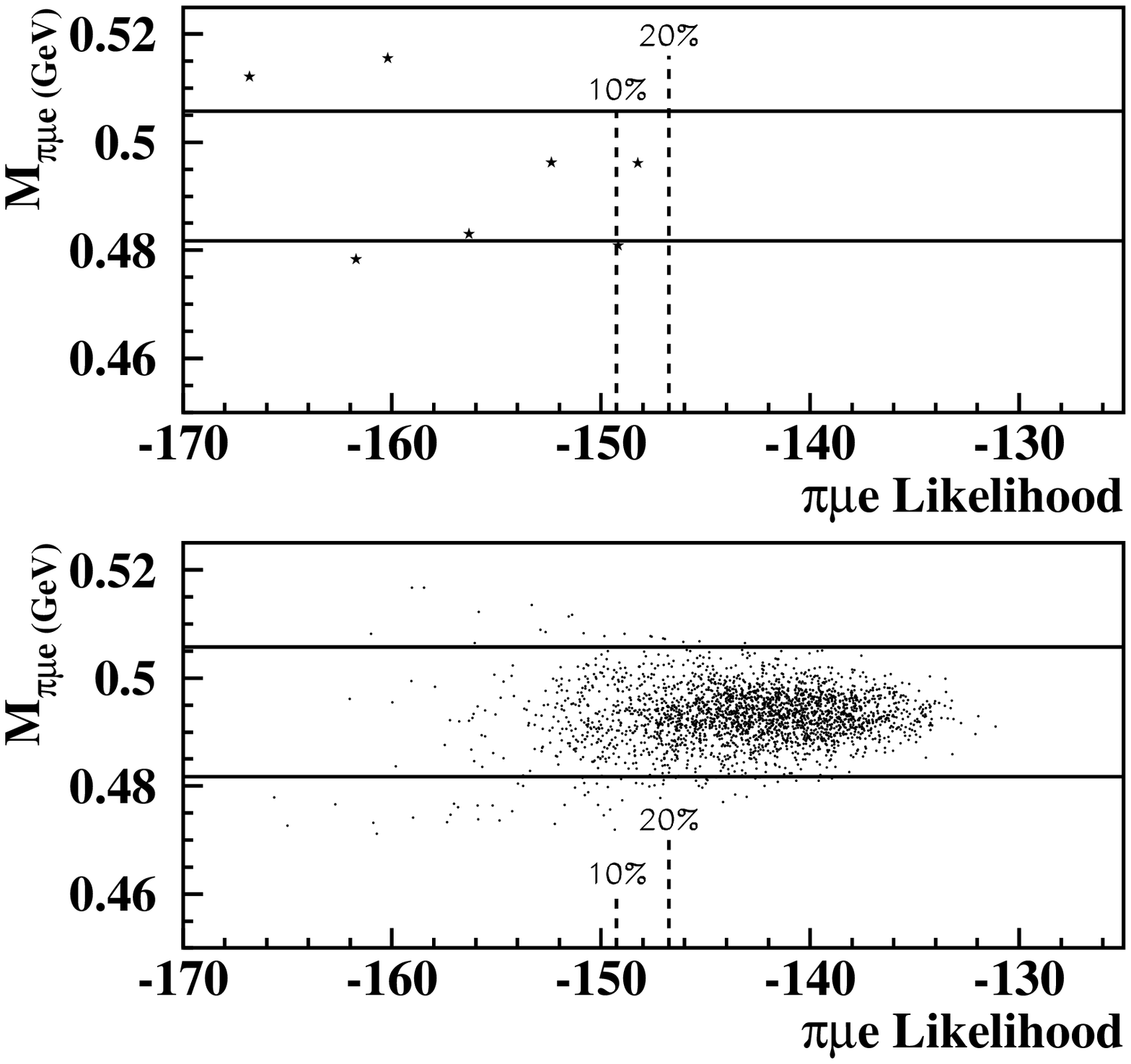}\centering\\
\includegraphics[width=59mm]{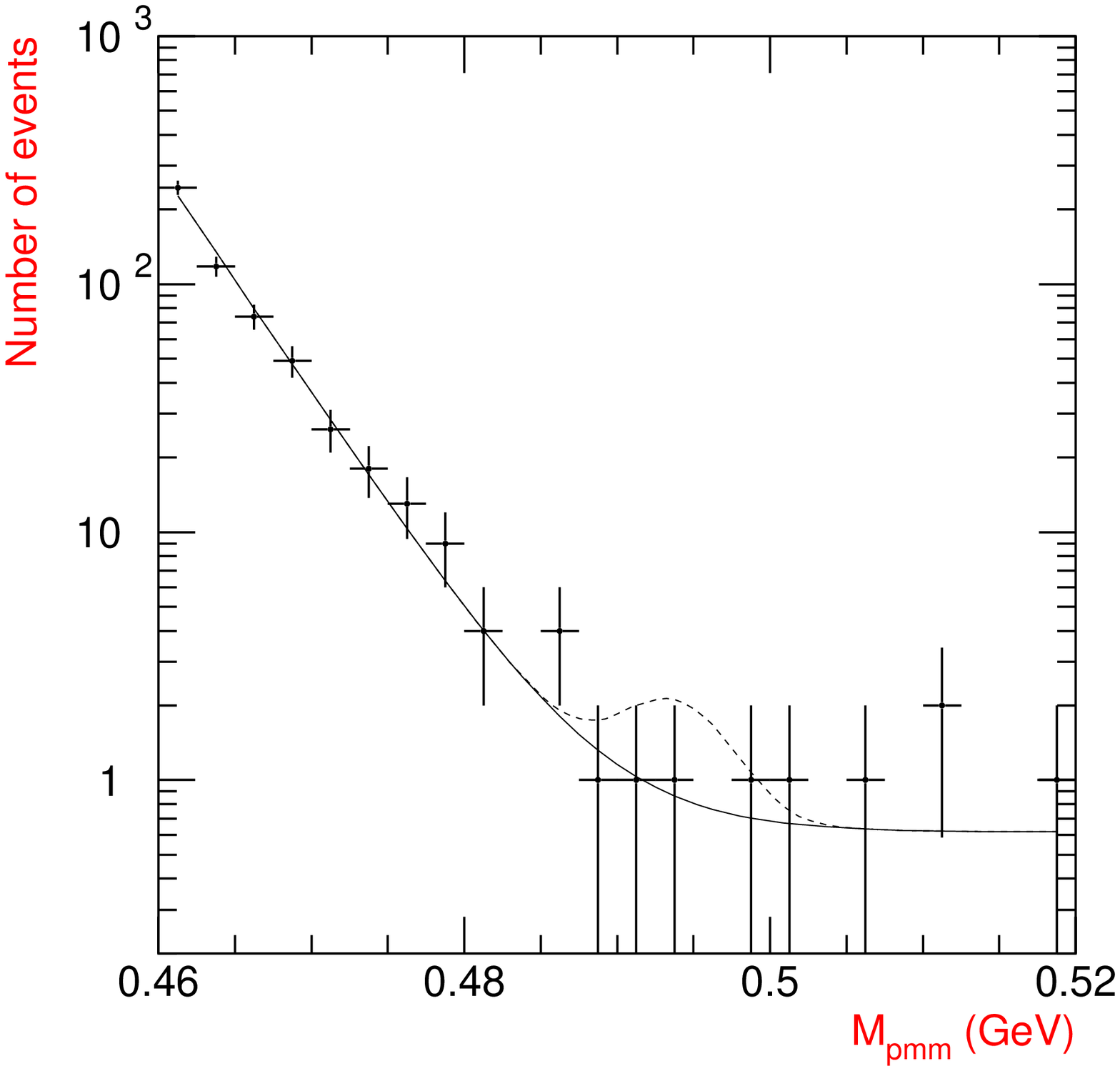}\hfill
\includegraphics[width=59mm]{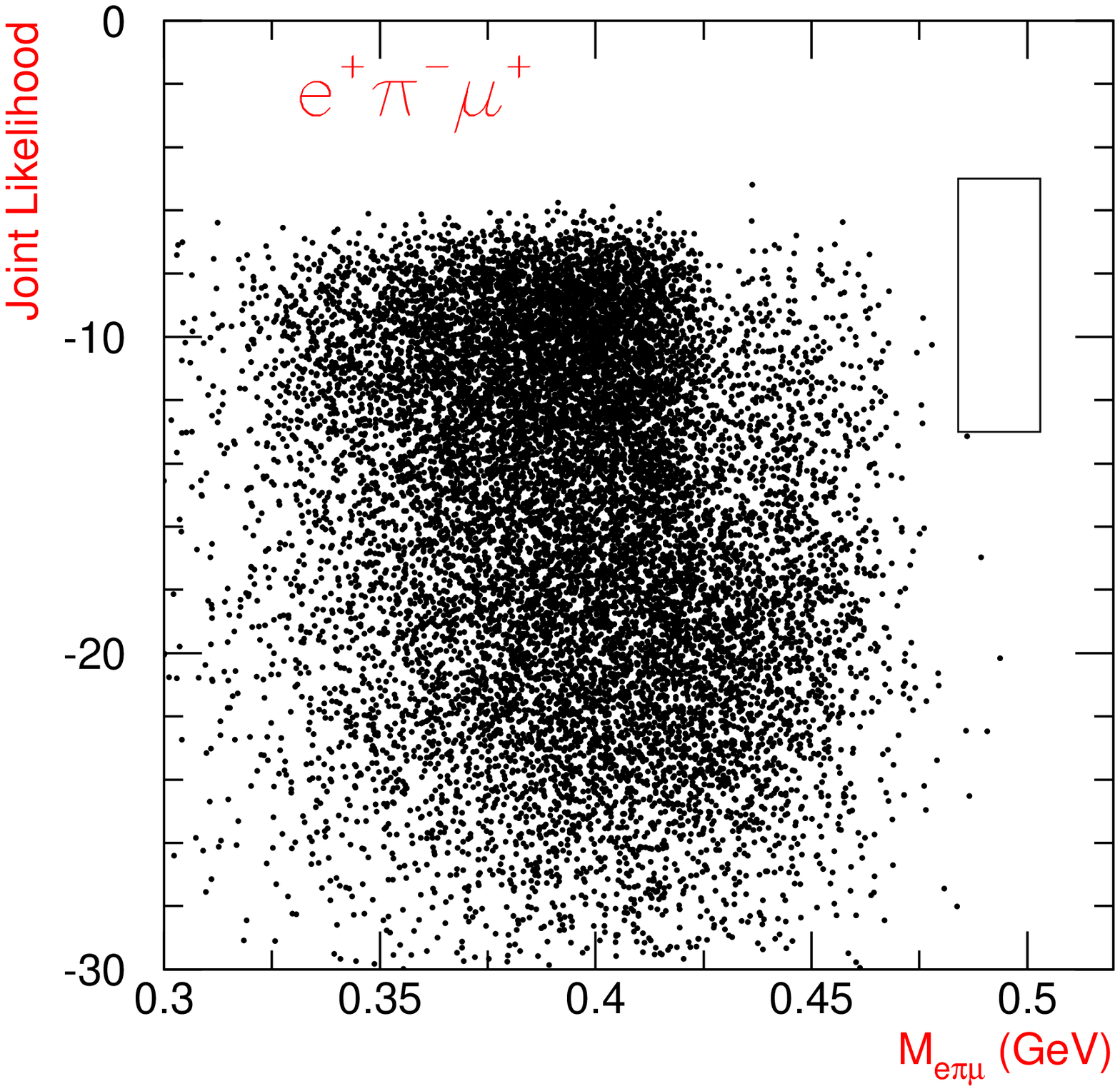}
\caption{\it Top: scatter plot of log-likelihood versus
invariant mass for $K_{\pi\mu e}$ candidate and Monte Carlo
generated signal events; the horizontal lines demark the
$3\sigma$ mass region\cite{Appel00a}. 
Bottom right: as above with coordinates
exchanged for
$K_{e\mu\pi}$ candidate events; the box indicates the signal region.
Bottom left: invariant mass distribution for $K_{\mu\mu\pi}$ candidate
events; the dashed line indicates the expected signal\cite{Appel00b}.
\label{fig:lfv}}
\end{figure}

Not all our $K_{\pi\mu e}$ data have been analysed yet. The data
taken during the last running period in 1998 are projected to push the
sensitivity limit below the $10^{-11}$ branching ratio level. 
Table~\ref{tab:lfv} lists the limits deduced from the 1995\cite{Pislak97} 
and 1996 data\cite{Appel00a}. Figure~\ref{fig:lfv} shows 
the mass distribution of the few $K_{\pi\mu e}$ candidates
remaining after a likelihood analysis. No event is found with a likelihood 
for the $K_{\pi\mu e}$ hypothesis exceeding 20\%. Background events may come
from $K^+$ decays with errors in kinematic reconstruction or particle
identification measurements or from accidental combinations of three tracks
(dominant contribution).
The probabilities to misidentify a $\pi^-$ as an $e^-$, or an a $\pi^+$
as an $e^+\;(\mu^+)$ are $2.6\times 10^{-6}$ and $1.7\times 10^{-5}\;(0.049)$,
while the identification efficiencies for $\pi^+,\;\mu^+$
and $e^-$ are 78\%, 74\% and 55\%, respectively.
The likelihood distributions constructed for these background events
let us then expect 2.6 events below the 20\% line indicated in 
Fig.~\ref{fig:lfv} consistent with the threeevents observed. 
Combined with earlier data\cite{Lee90} our results yield an upper limit for the
$K_{\pi\mu e}$ branching ratio of $2.8\times 10^{-11}$ (see 
Table~\ref{tab:lfv}).

In models described by a horizontal gauge 
interaction\cite{Cahn80,Shanker81} 
one may use this branching ratio to deduce
a limit for the mass of an intermediate
boson $X$ by comparison to the allowed decay  
$K^+\rightarrow \pi^0 \mu^+ \nu_\mu$ ($K_{\mu3}$):
\[ \frac{B.R.(K_{\pi \mu e})}{B.R.(K_{\mu 3})}\approx \frac{16}{\sin^2\theta_c}
\left (\frac{g_X}{g}\right )^4\left (\frac{M_W}{M_X}\right )^4
\;\Rightarrow\;(g/g_X)M_X < 60\ {\rm TeV}/c^2\ .\]
Here $\theta_c$ is the Cabbibo angle,
and $g_X$ and $g$ are the couplings for the new and the weak interaction, 
respectively.

Unlike  $K_{\pi\mu e}$, which only violates lepton flavour conservation, 
the other decays for which Table~\ref{tab:lfv} lists upper limits,
also violate generation number conservation\cite{Cahn80}. We used our
$K_{\pi\mu\mu}$ and $K_{e4}$ data samples to search for these decays. 
These data sets were taken with dedicated triggers at lower beam
intensity. The $K_{\mu\mu\pi}$ and
$K_{e\mu\pi}$ mass plots are shown in Fig.~\ref{fig:lfv} as
examples. The most significant reduction of the existing limit\cite{Caso00}
(by a factor of 50000) was obtained for $K_{\mu\mu\pi}$, a decay which is
sensitive to the same exotic mechanism, which may permit nuclear
double $\beta$-decay\cite{Littenberg00}.

\section{$K_{e4}$ and the $\pi\pi$ scattering length}
\label{sec:ke4}

Since more than 30 years, the $K_{e4}$ decay has been used to 
study $\pi\pi$-scattering at low energies. 
The most recent experiment performed by a
Geneva-Saclay collaboration at CERN~\cite{Rosselet77} accumulated
30000 events.
Using these data and some peripheral $\pi N\rightarrow \pi\pi N$
data did lead to the presently 
generally accepted value for the isoscalar $s$-wave
scattering length of $a_0^0=(0.26\pm0.05)$ 
(in units of $M_{\pi}^{-1}$). 

On the theoretical side $\pi\pi$ scattering is the
{\em golden reaction} for chiral perturbation theory (ChPT)~\cite{Colangelo99}.
This low-energy effective theory for the strong interaction
makes firm predictions for the scattering length.  
The pioneering tree level calculation by Weinberg~\cite{Weinberg66} yielded
$a_0^0=0.156$ $(a_0^2=-0.046)$. The one-loop 
($a_0^0=0.201$, $a_0^2=-0.042$~\cite{Gasser83}) and 
the recently completed 
two loop calculation ($a_0^0=0.217$, 
$a_0^2=-0.041$~\cite{Bijnens96,Bijnens98}) 
show a satisfactory 
convergence. The most recent calculation~\cite{Colangelo00} 
matches the known chiral perturbation theory representation 
of the $\pi\pi$ scattering amplitude to two loops 
with a phenomenological description that relies on the Roy equations
(see below). The corrections to Weinberg's low energy theorems 
for the $s$-wave scattering lengths are worked
out to second order in the expansion in powers of the quark masses,
and the uncertainties from higher order effects are estimated
reliably for the first time. 
The resulting predictions $a_0^0=0.220 \pm 0.005$, 
and $a^2_0=-0.0444 \pm 0.0010$ are in remarkable agreement
with our new experimental result, which we will 
discuss in the following.

\begin{figure}[htb]
\includegraphics[width=63mm]{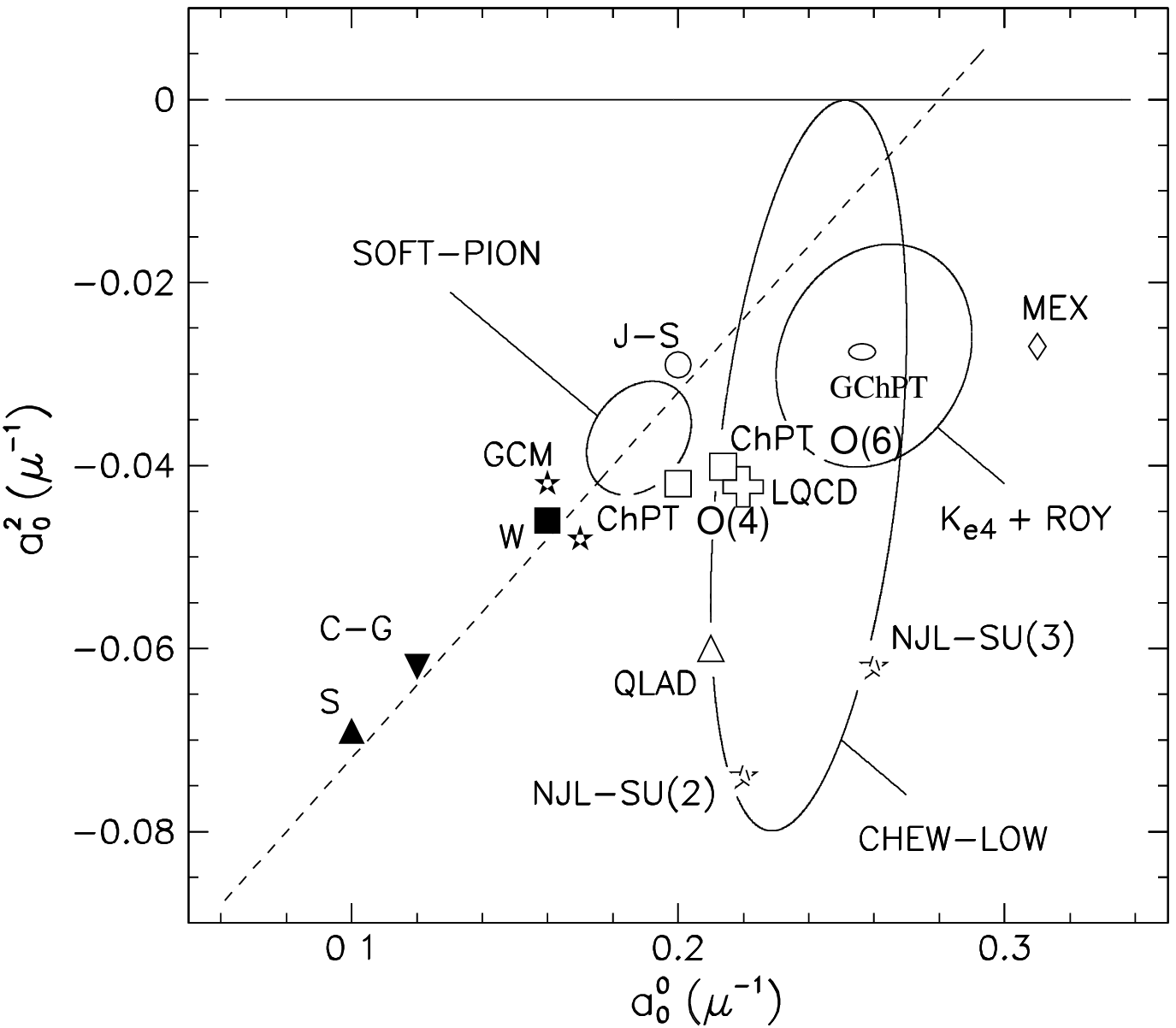}\hfill
\includegraphics[width=53mm]{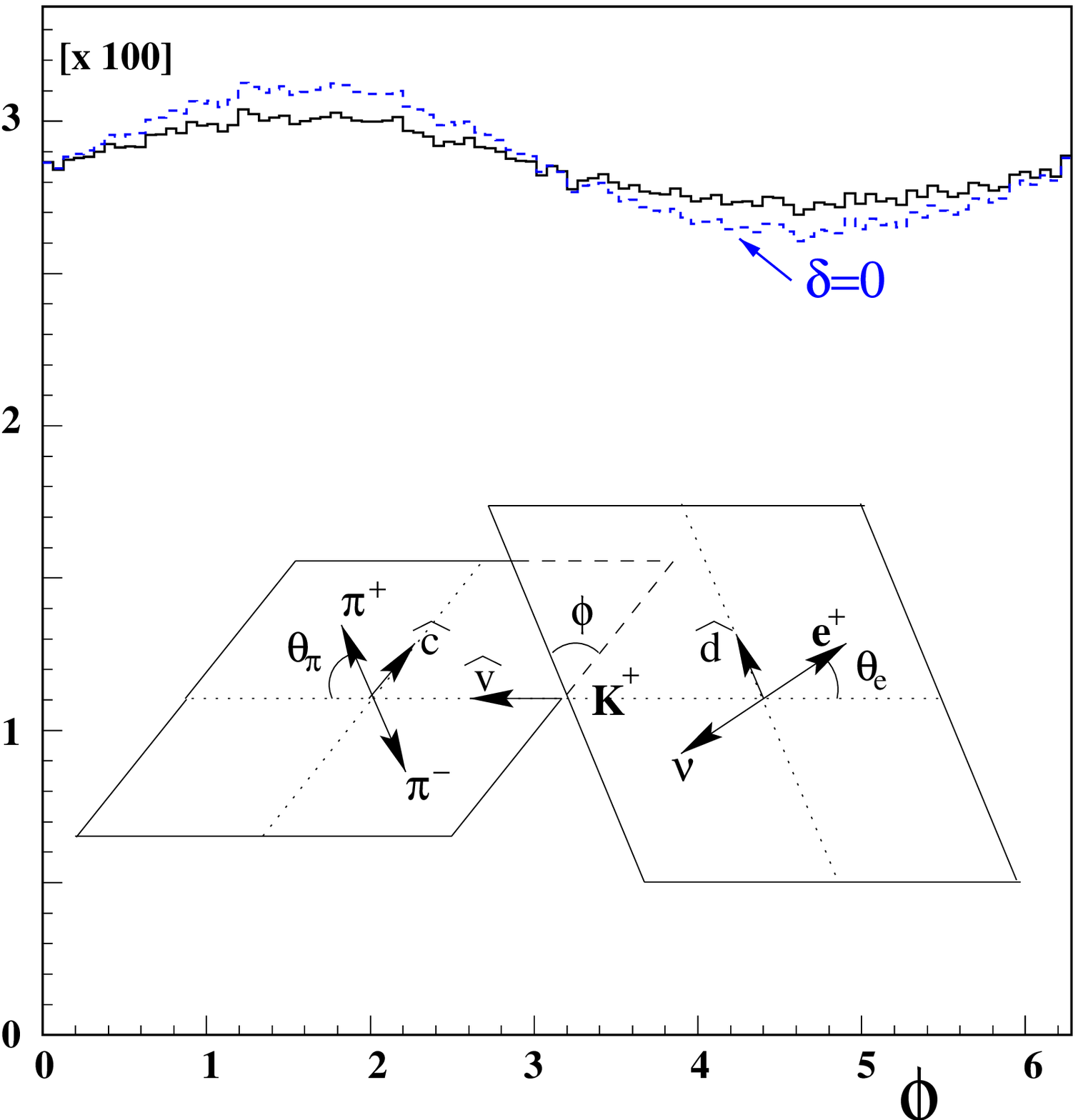}\hfill

\caption{\it Left: previous measurements and predictions\cite{Pipith} for
the scattering length (taken from Ref.\cite{Pocanic94}). 
Right: azimuthal asymmetry averaged
over all other kinematical quantities with standard phase
$\pi\pi$ shifts and with phase shifts set to zero.
Insert: kinematical quantities used in the description
of $K_{e4}$ decay.\label{fig:pipi1}}
\end{figure}

Figure~\ref{fig:pipi1} shows these predictions and also
a number of older, alternative model theoretical predictions\cite{Pipith}.
Since the extraction of the $\pi\pi$ amplitude in the high statistics
$\pi N \rightarrow \pi\pi N$ experiments is model dependent,
and the $\pi\pi$ atom data from the DIRAC experiment\cite{Schacher00}
are not yet available,
one had to rely primarily on the $K_{e4}$ data
for the comparison between theory and experiment, which were
not sufficiently accurate to discriminate between the
different models.

$K_{e4}$ events are selected with a vertex within the decay
tank of acceptable quality, a momentum
reconstructed from the three daughter particles below the
beam momentum. The momentum vector 
of the incoming $K^+$ is obtained from the hit in the pixel 
counter (see Fig.~\ref{fig:det}) and the vertex, and requiring 
the $K^+$ to come from the production target 27.5~m upstream 
as well as an unambiguous identification of the $e^+$ and the
$\pi^-$ in both \v{C}erenkov counters and the calorimeter.
The event sample contains a small amount of background events,
mainly from $K_{\tau}$ decay ($1.3\pm0.3$ \%) and accidentals
($2.6\pm0.2$ \%). 
A $K_{\tau}$ can fake a $K_{e4}$ by either a
misidentification of one of the $\pi^+$ due to $\delta$-rays, 
photomuliplier noise or the presence of an additional 
positron, or a decay of one $\pi^+$ directly or via a 
$\mu^+$ into a $e^+$. The dominating accidental background is a
combination of a $\pi^+\pi^-$ pair from $K_{\tau}$ decay with a 
positron from either the beam or from a $K_{dal}$ decay.

After event selection 406'103 events remain, of which we estimate 
388270 $\pm$ 5025 to be  $K_{e4}$ events. This corresponds to an increase in
statistics by more than a factor 12 compared with previous 
experiments~\cite{Rosselet77}.

\begin{figure}[htb]
\vspace*{-10mm}
\includegraphics[width=58mm]{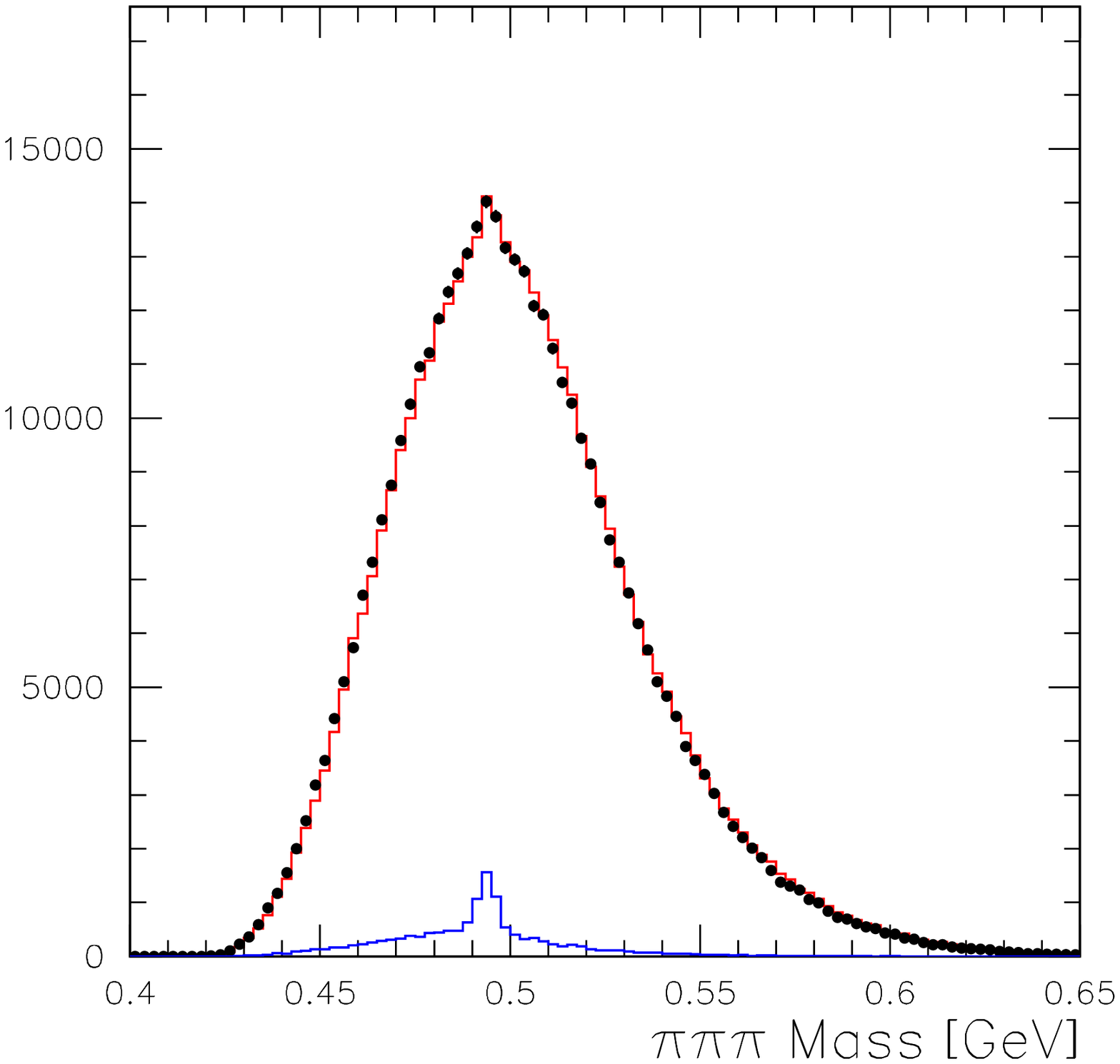}\hfill
\includegraphics[width=58mm]{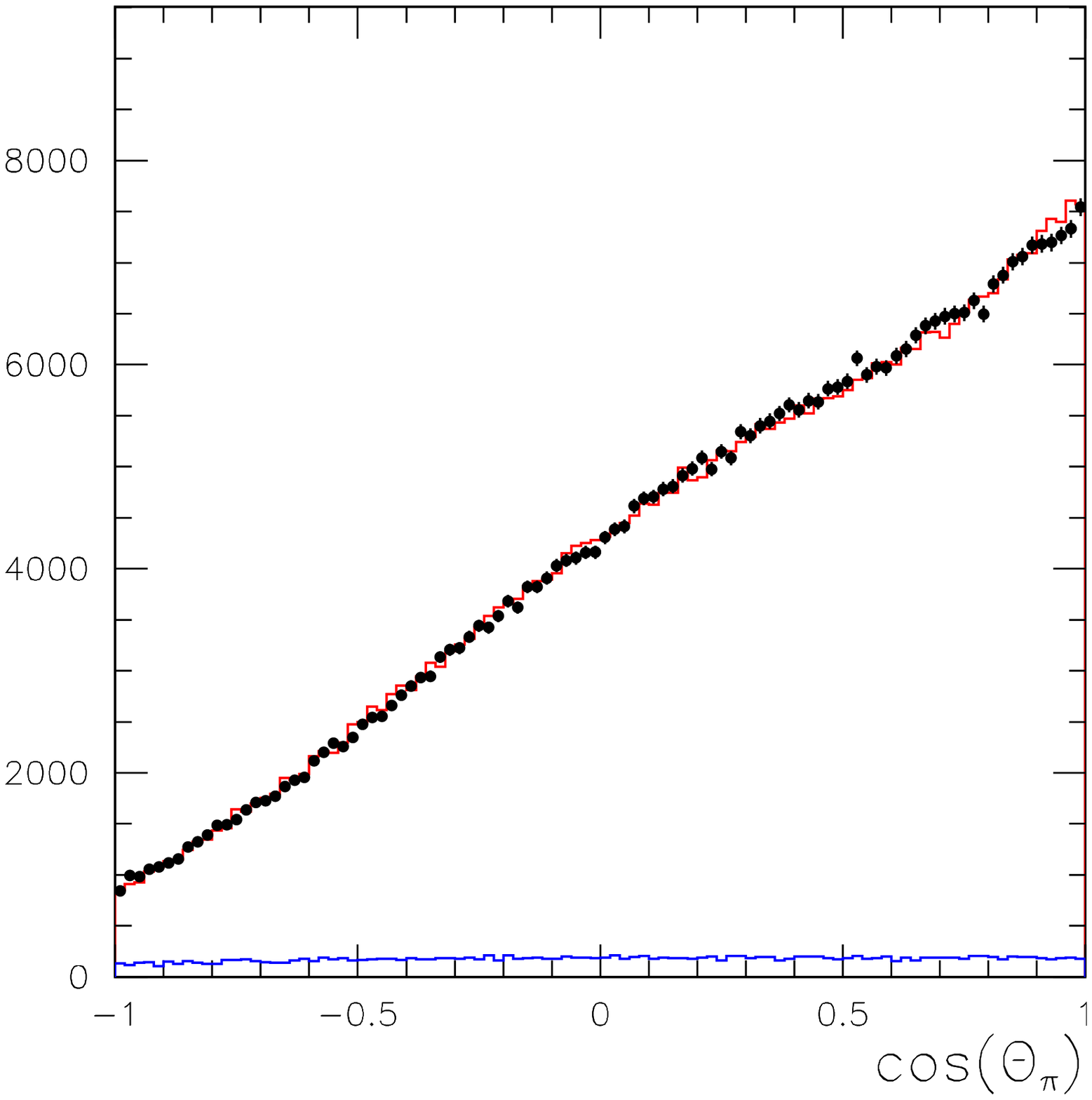}\\
\includegraphics[width=58mm]{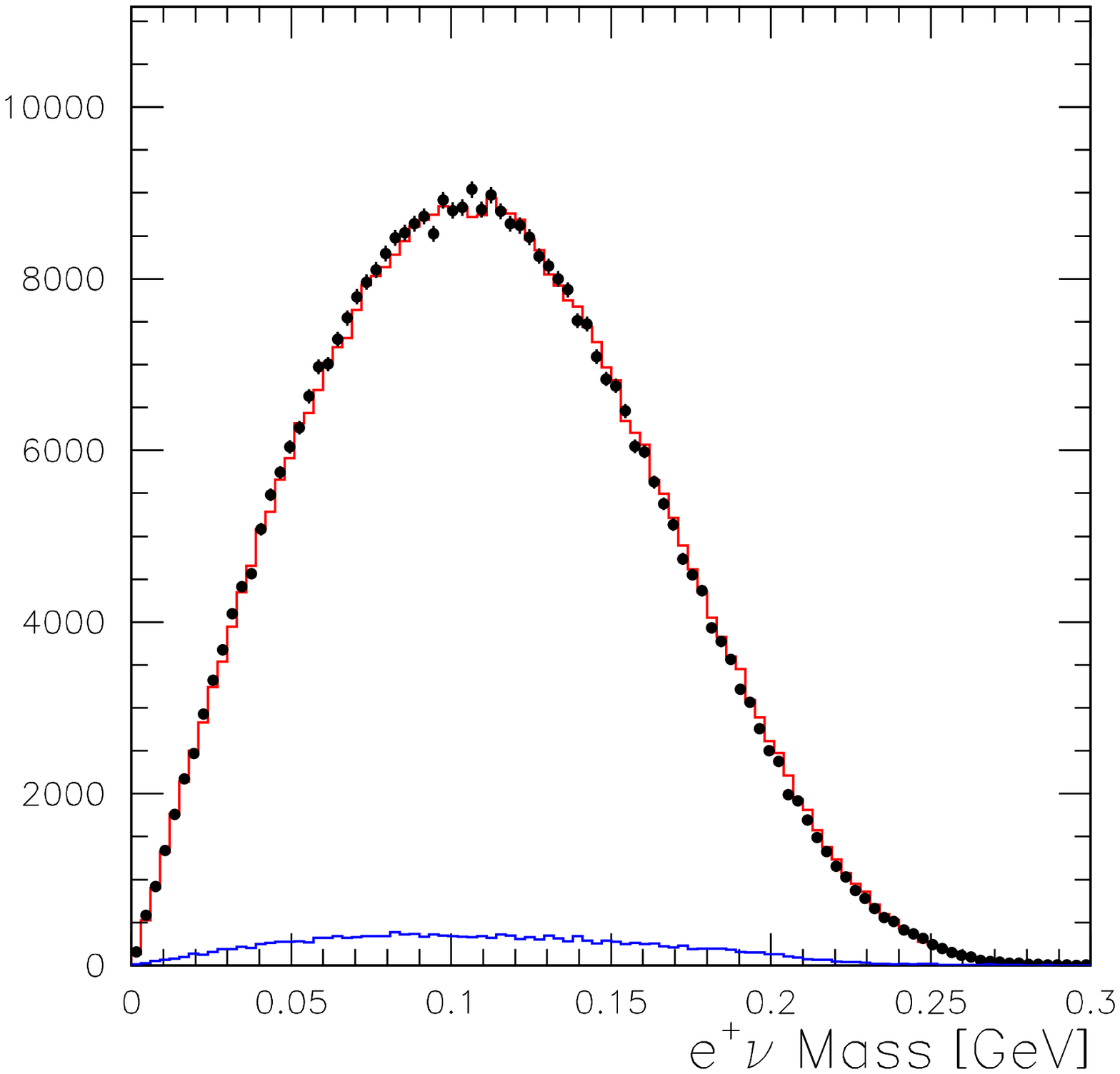}\hfill
\includegraphics[width=58mm]{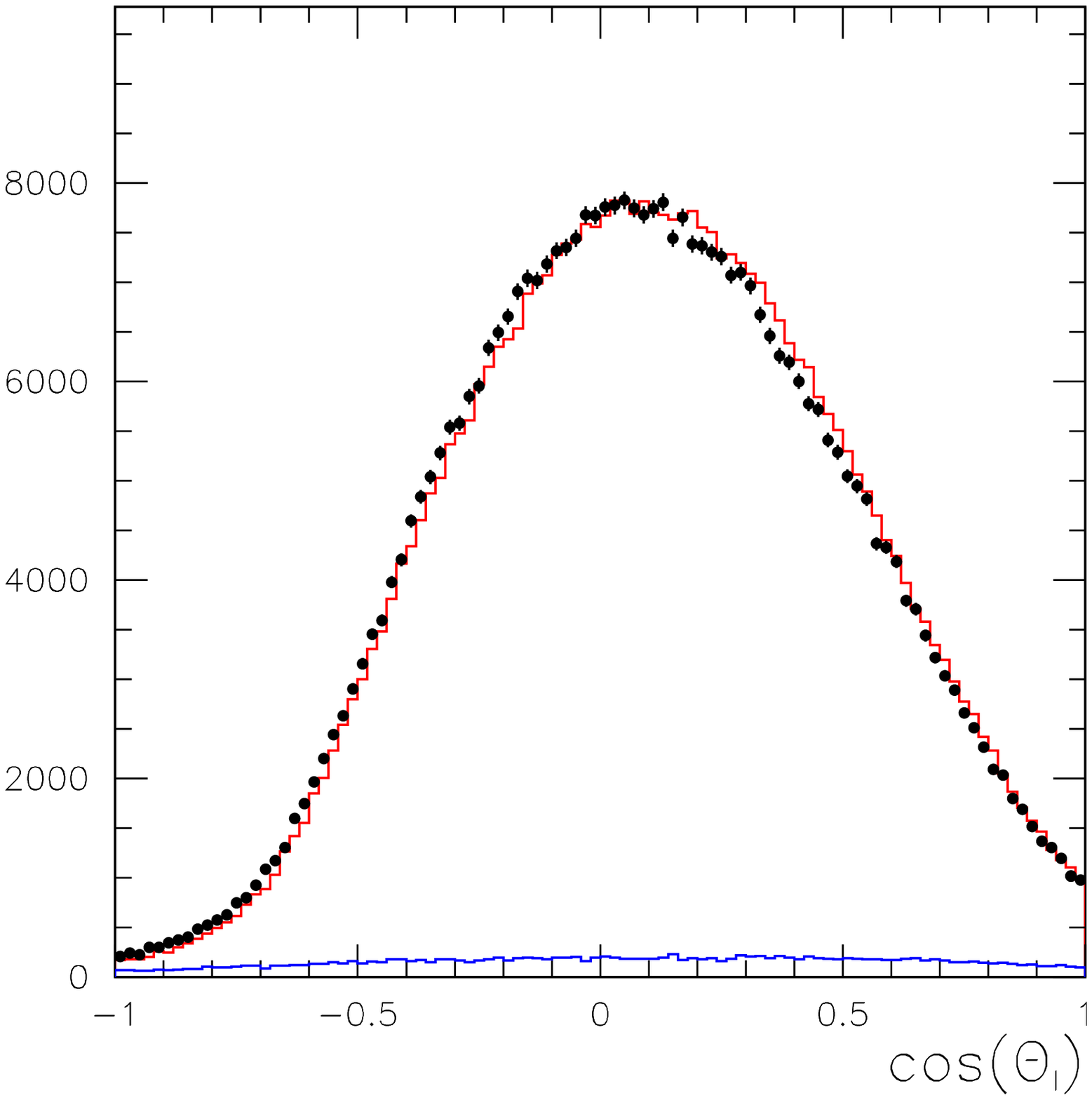}\\
\includegraphics[width=58mm]{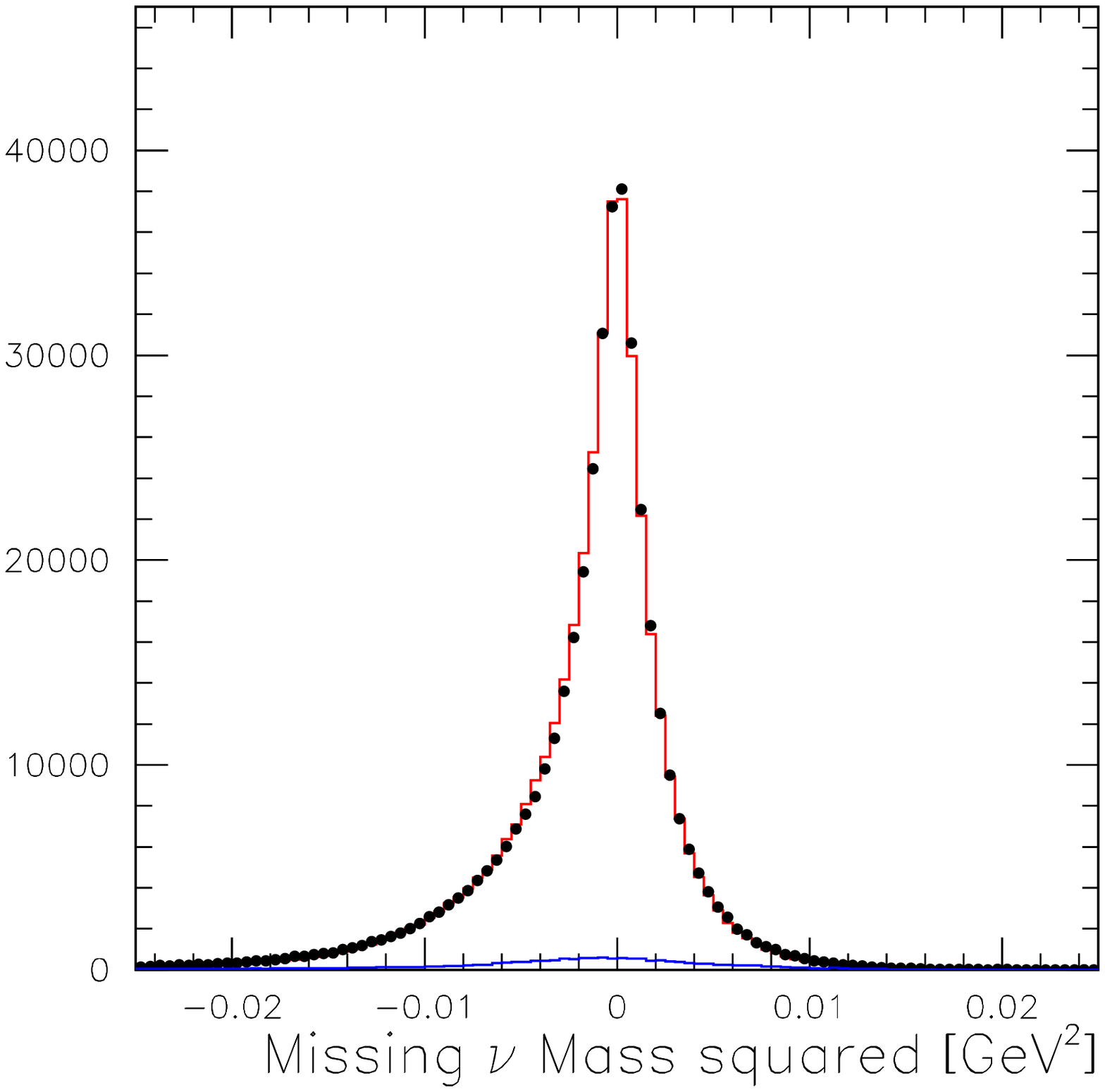}\hfill
\includegraphics[width=58mm]{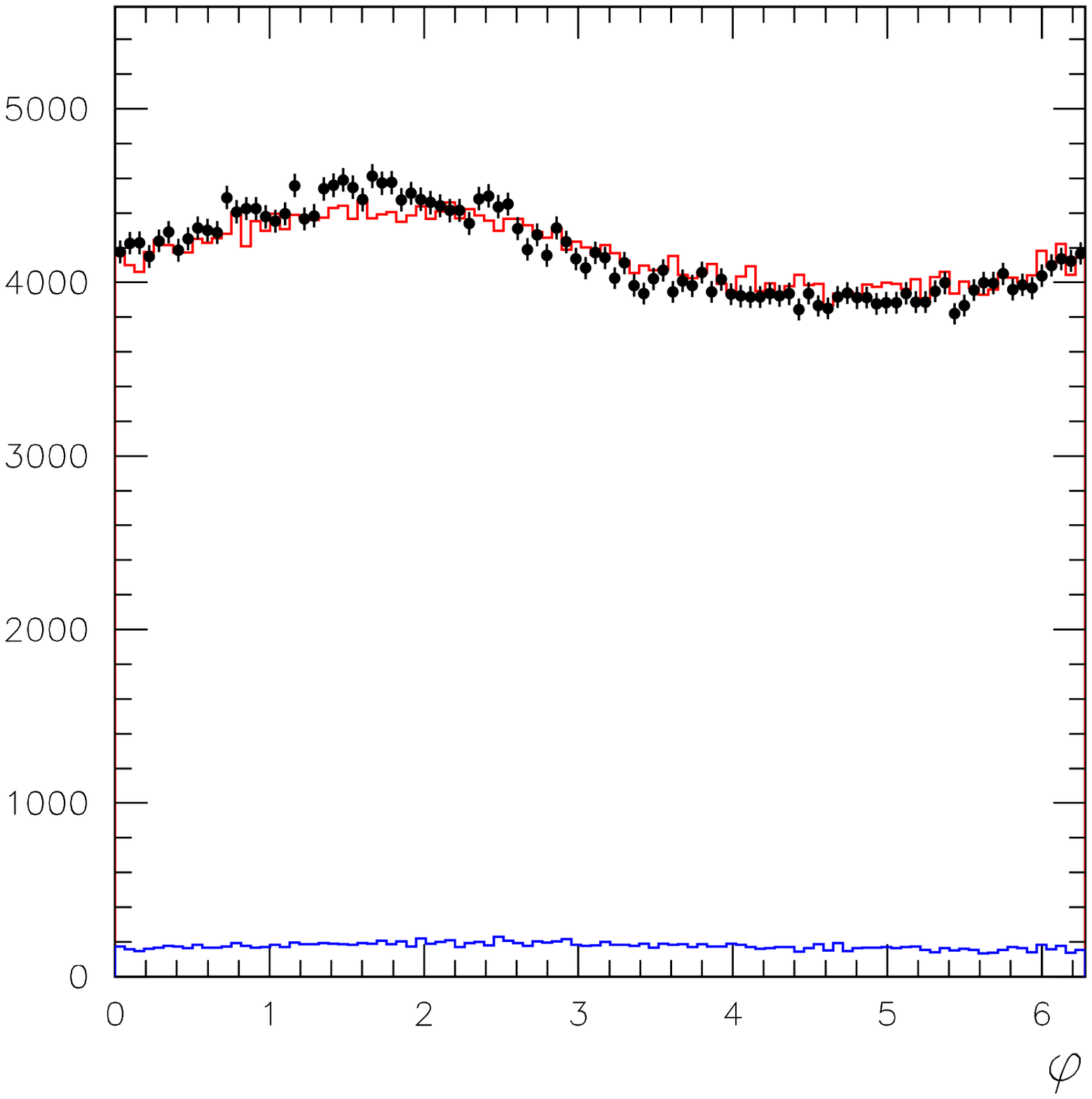}
\caption{\it $K_{e4}$ control plots. From top left to bottom right:
$\pi\pi$ mass; $\cos\theta_\pi$; $e\nu$ mass; $\cos\theta_e$;
missing $\nu$ mass squared; $\phi$. Data (solid circles), 
Monte Carlo (histogram) including backgrund, and background
separately (lower histogram).
\label{fig:pipi2}}
\end{figure}

Quantitative understanding and analysis of the
data relies strongly on the quality of Monte Carlo simulation
of the detector response. This GEANT-based\cite{Pislak97} simulation 
is supplemented with separately measured efficiencies of the
various detector parts, and was extensively tested against
data in the various decay channel under study. Figure~\ref{fig:pipi2}
shows a few control plots demonstrating the good agreement
between measured and simulated distributions of the kinematic variables.
The total Monte Carlo data sample comprises 81.6$\cdot 10^6$
generated  $K_{e4}$ events of which 2.9$\cdot 10^6$ were accepted after
reconstruction and passage through the same analysis codes as the data.
We also compare Monte Carlo with data distributions for the 
kinematically very distinct $K_{\tau}$ and $K_{dal}$ decays, and
are able to reproduce the measured branching ratios for these decays.
A chiral perturbation theory calculation on the one loop
level~\cite{Bijnens90,Bijnens92,Riggenbach91} is used to model
the physics of the decay at this point. Radiative corrections 
are included following Diamant~\cite{Diamant76a}.

For the description of the $K_{e4}$ decay
three reference frames are commonly used (see Fig.~\ref{fig:pipi1}): 
the $K^+$ rest system ($\Sigma_K$), the $\pi^+\pi^-$ rest system
($\Sigma_{\pi\pi}$) and the $e^+\nu$ rest system ($\Sigma_{e\nu}$). 
The decay kinematic is then fully described by five 
variables: $s_\pi=M_{\pi\pi}^2$, the invariant mass squared of the dipion;
$s_e=M_{e\nu}^2$, the invariant mass squared of the dilepton; 
$\theta_\pi$, the angle of the $\pi^+$ in $\Sigma_{\pi\pi}$
with respect to the direction of flight of the dipion in
$\Sigma_K$; $\theta_e$, the angle of the $e^+$ in $\Sigma_{e\nu}$
with respect to the direction of flight of the dilepton in
$\Sigma_K$; $\phi$, the angle between the plane formed by the pions in
$\Sigma_{\pi\pi}$ and the corresponding plane formed by the
leptons.
The experimental resolution for these five variables is
$\sigma(s_{\pi})=$ 0.00133~GeV$^2$, $\sigma(s_{e})=$ 0.00361~GeV$^2$,
$\sigma(\theta_{\pi})=$ 0.147, $\sigma(\theta_{e})=$ 0.111 and
$\sigma(\phi)=$ 0.404.

The $K_{e4}$ branching ratio is determined relative to $K_{\tau}$ 
decay (Br$(\tau)=5.59\pm0.05$ \%), 
for which events were collected with a minmum bias
trigger concurrently with $K_{e4}$ events. 
We obtain 
\[ {\rm BR}(K_{e4})= \left(4.109\;\pm\;0.008\;({\rm stat.})
\pm 0.110\;({\rm syst.})\right)\cdot 10 ^{-5} \]
The result is in agreement with the average of previous measurements: 
$(3.91 \pm 0.17)\cdot 10^{-5}$~\cite{Caso00}.
The largest of the 17 contributions to the systematic error (2.68\%),
which we have evaluated, stem from the background subtraction (1.2\%),
the \v{C}erenkov efficiencies (1.5\%) and the
$K_{\tau}$ branching ratio (0.9\%).

The analysis of the $K_{e4}$ decay distributions
is far from trivial, and requires a fair amount
of theoretical input. We list below the basic ingredients and
prepare the notation for the subsequent result section.

The most general form of the decay matrixelement 
is written most conveniently using the following combinations
of four-momenta and Lorentz invariants:
\begin{equation}\begin{array}{c}
P=p_1+p_2,  \; Q=p_1-p_2, \; L=p_e+p_\nu,\; 
X=[(P\cdot L)^2-s_\pi s_e]^{1/2}\ ,\nonumber\\[1ex]
M=\frac{\textstyle G_F}{\textstyle \sqrt{2}} V^*_{us} 
\overline{u}(p_{\nu})\gamma_{\mu}
(1-\gamma_5)v(p_e)(V^{\mu}-A^{\mu})\ ,\nonumber\\[1ex]
A^{\mu}=\frac{\textstyle 1}{\textstyle M_K}
\left(F P^{\mu}+G Q^{\mu} + R L^{\mu}\right)\ ,
\hspace*{5mm}
V^{\mu}=\frac{\textstyle H}{\textstyle M_K^3}
\epsilon^{\mu\nu\rho\sigma}L_{\nu}P_{\rho}Q_{\sigma}\ .
 \end{array}\nonumber
\label{eq:Matrix}
\end{equation}
$F$, $G$, $R$, $H$ dimensionless complex functions of
 $p_1\cdot p_2$, $p_1 \cdot p$, $p_2 \cdot p$,
or of $s_\pi$, $s_e$ and $\theta_\pi$.
If terms $\propto M_e^2/s_e$ are neglected $R$ does not
contribute to the decay rate, first evaluated
by Pais and Treiman\cite{Pais67}:
\begin{equation}
d\Gamma_5 = \frac{G_F^2 V^2_{us}}{2^{12}\pi^6M_k^5} X
              \frac{2M_\pi q}{\sqrt{s_\pi}} 
J_5(s_\pi,s_e,\theta_\pi,\theta_e,\phi)
              ds_\pi ds_e d\cos\theta_\pi d\cos\theta_e d\phi\ .
\label{eq:Pais}\end{equation}
For the detailed decomposition of the factor $J_5$
in the above expression in terms
of the kinematical variables we must refer to Ref.\cite{Pais67,Amoros99}.
A convenient partial wave expansion of the form factors 
in the variable $\theta_{\pi}$ gives the following~\cite{Amoros99}:
\begin{equation}\begin{array}{lcl}
F&=&\left(f_s+f_s^\prime \,q^2+f_s^{\prime\prime}\, q^4+f_e \, s_e\right)
    \exp(i\delta^0_0(s_\pi))\\
 &&\hspace*{20mm}+\tilde{f}_p \,(q^2/s_\pi)^{1/2} \,(P\cdot L)\,\cos\theta_\pi \,
    \exp(i\delta^1_1(s_\pi))\ ,\nonumber \\[1ex]
G&=&\left(g_p + g_p^\prime \,q^2 + g_e \,s_e \right) \exp(i\delta^1_1(s_\pi))
\ ,\nonumber\\[1ex]
H&=&\left(h_p+h_p^\prime \,q^2\right) \exp(i\delta^1_1(s_\pi))\ ,\end{array} 
\label{eq:AB99}
\end{equation}
where $q$ is the pion momentum in $\Sigma_{\pi\pi}$, divided by $M_\pi$.
This parameterization yields ten new form factors $f_s$, $f_s^\prime$, 
$f_s^{\prime\prime}$, $f_e$, $\tilde{f}_p$, $g_p$, $g_p^\prime$, $g_e$,
$h_p$, and $h_p^\prime$, which do not depend on any kinematical
variables, plus the phases $\delta_0^0$ and $\delta_1^1$,
which are still functions of $s_\pi$. 

It is worth noting at this point, that the phase shifts enter into the decay
rate only through interference terms,
which are proportional to $\sin\phi$ or $\cos\phi$. The azimuthal $\phi$
asymmetry, which has an amplitude of only approximately 10\% of the
the average (Fig.~\ref{fig:pipi2}), is principal source of the
phase shift information as illustrated in Fig.~\ref{fig:pipi1}.

There exist three slightly different possibilities to extract the 
form factor and phase shift information from the
data. Firstly, we can use the parameterization given above
in Eq.~\ref{eq:AB99}
and perform individual fits in bins of $s_\pi$ respecting the
$s_\pi$ dependence of the phase shifts. Secondly,
there is the option to express the phase shifts by 
Eq.~\ref{eq:schenk} shown below. This allows to use the 
whole data sample in one single fit. Finally, we can employ in addition
Eq.~\ref{eq:UB}, reducing the number of parameters
by one. 

The phase shifts  $\delta_0^0$ and $\delta_1^1$
can be related to  the $s$-wave 
scattering lengths $a_0^0$ and $a_0^2$
by means of the Roy equations~\cite{Roy71}. A recent
analysis of the phase shift data~\cite{Ananthanarayan00}
used the parametrisation proposed by Schenk~\cite{Schenk91}:
\begin{equation} 
\tan \delta_\ell^I =\sqrt{1-{4 M_\pi^2 \over s_\pi}}\; 
q^{2 \ell} \left\{A^I_\ell
+ B^I_\ell q^2 + C^I_\ell q^4 + D^I_\ell q^6 \right\}\left({4
  M_\pi^2 - s^I_\ell \over s_\pi-s^I_\ell} \right) \ .
\label{eq:schenk}
\end{equation} 
The Roy equations were solved numerically, expressing the parameters
$A^I_\ell$, $B^I_\ell$, $C^I_\ell$, $D^I_\ell$,
and $s^I_\ell$ as a function of the scattering lengths
$a_0^0$ and $a_0^2$. The possible values of
the scattering lenghts were restricted to a band in the 
$a_0^0-a_0^2$ plane, the {\em universal band} introduced
by Morgan and Shaw~\cite{Morgan69}.
The centroid of this band, the {\em universal curve}
can be expressed by a second
degree polynominal, relating $a_0^0$ and $a_0^2$:
\begin{equation} 
a_0^2 =-0.0849+0.232\, a_0^0-0.0865\, (a_0^0)^2   
\label{eq:UB}
\end{equation} 
 
For the fits our data are divided over
6 bins in $s_\pi$, 5 bins in $s_e$, 10 bins in 
$\cos\theta_\pi$, 6 bins in $\cos\theta_e$, and 16 bins in $\phi$, 
28'800 bins in total with on average 13 events in each
bin. In the $\chi^2$ minimization procedure the
number of measured events $n_j$ in each bin $j$
is compared to the number of expected events in that bin $r_j$
given by 
\begin{equation}
r_j =\frac{{\rm Br}(K_{e4})N(K^+)}{N(\mbox{Generated events})} 
          \times 
    \sum \frac{J_5(F^{New}(q),
           G^{New}(q),H^{New}(q))}
           {J_5(F^{MC}(q),G^{MC}(q),H^{MC}(q))}\ .
\label{eq:fit}
\end{equation}
$F^{MC}$, $G^{MC}$, and $H^{MC}$ correspond to the original form 
factors used to generate the 
Monte Carlo event, while $F^{New}$, $G^{New}$, and $H^{New}$ are
the form factors derived from the parameters of the fit.
$q$ denotes the inital and not the reconstructed kinematical
variables. Thus the parameters are applied on an event by event basis,
and at the same time a possible bias caused by the initial choice of
the matrix element is eliminated.

\begin{figure}[htb]
\includegraphics[width=59mm]{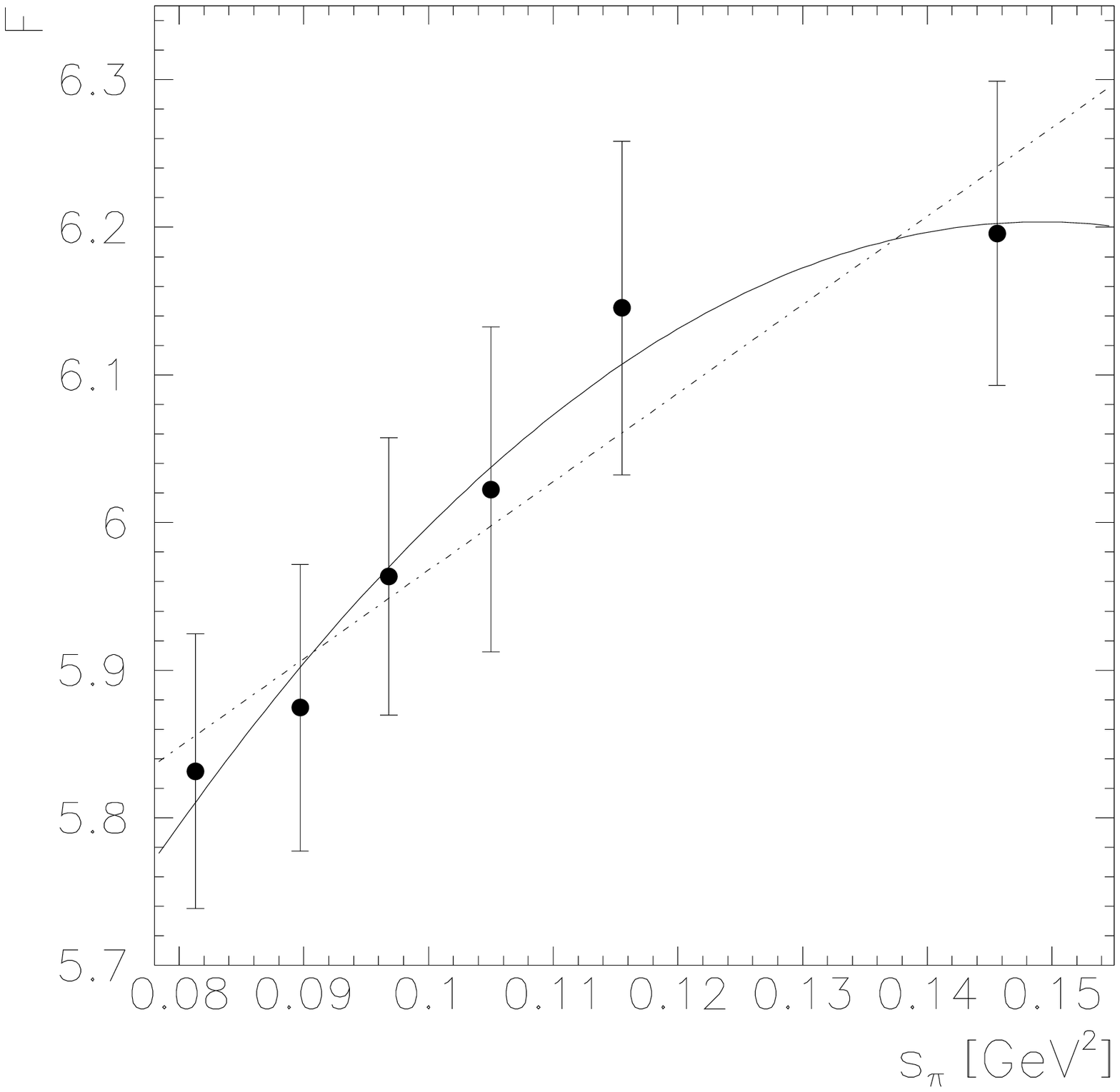}\hfill
\includegraphics[width=59mm]{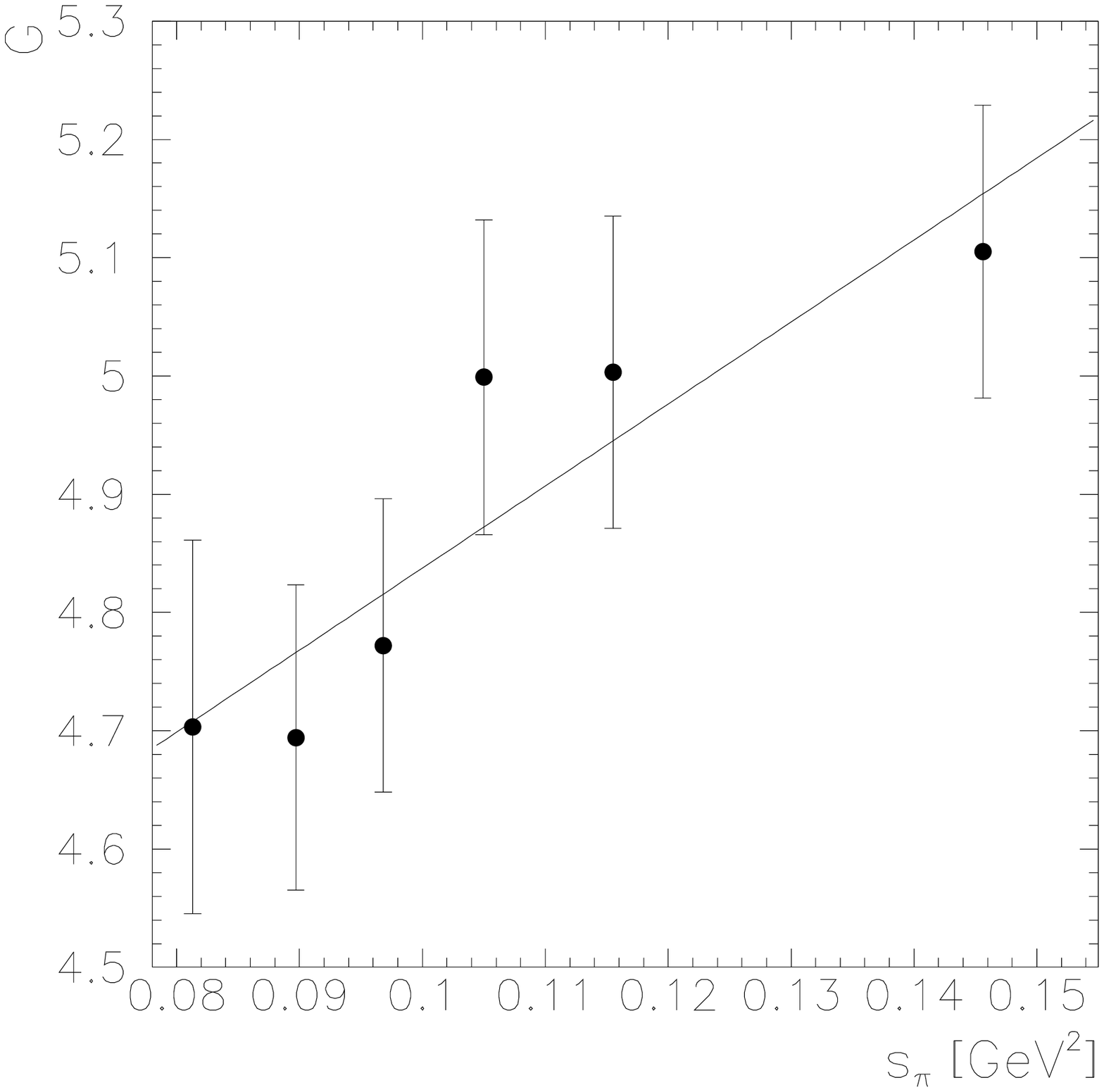}
\caption{\it Form factors $F$ and $G$ as function of $s_\pi$.
\label{fig:pipi3}}
\end{figure}

\begin{figure}[htb]
\includegraphics[width=60mm]{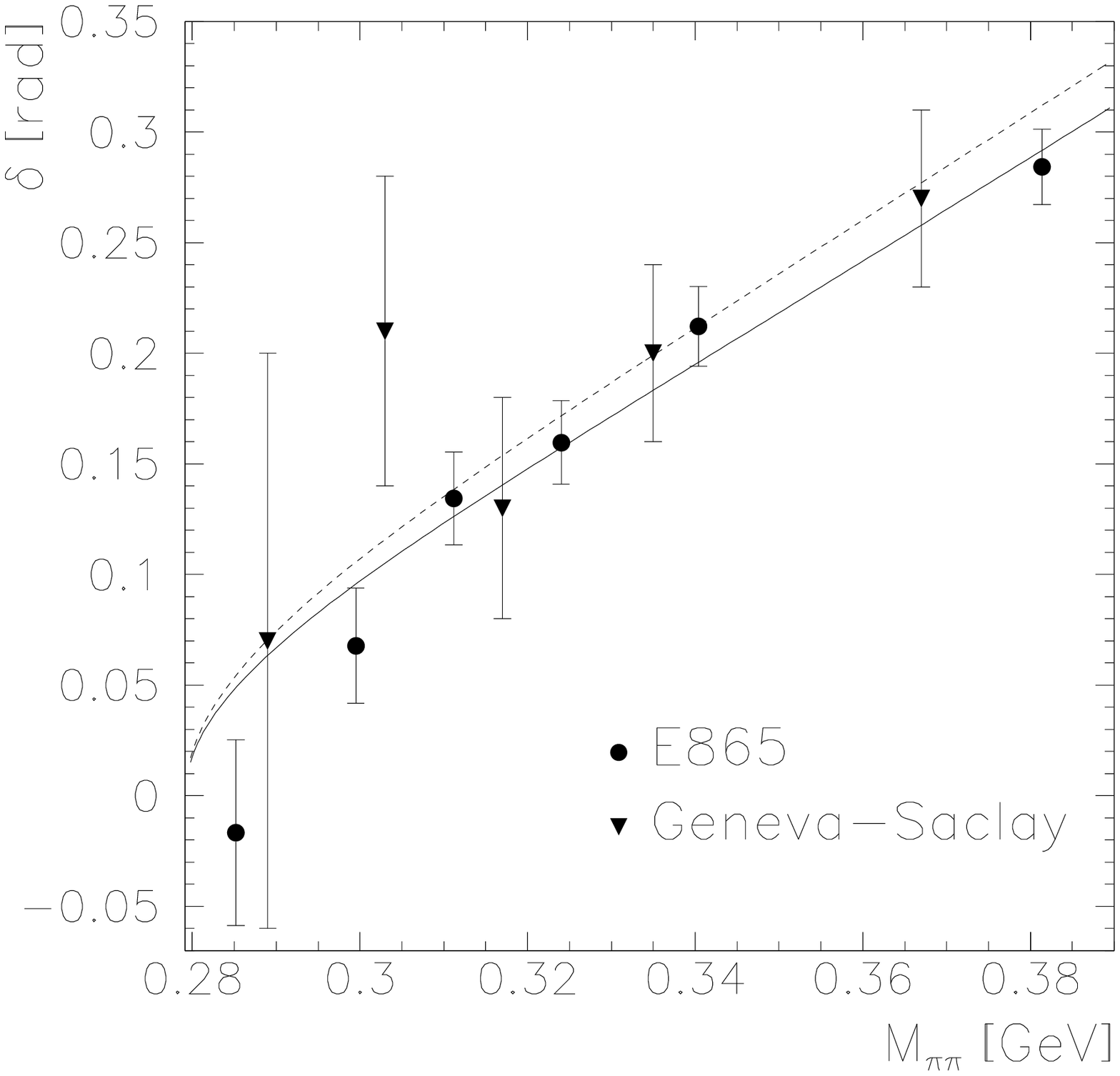}\hfill
\includegraphics[width=58mm]{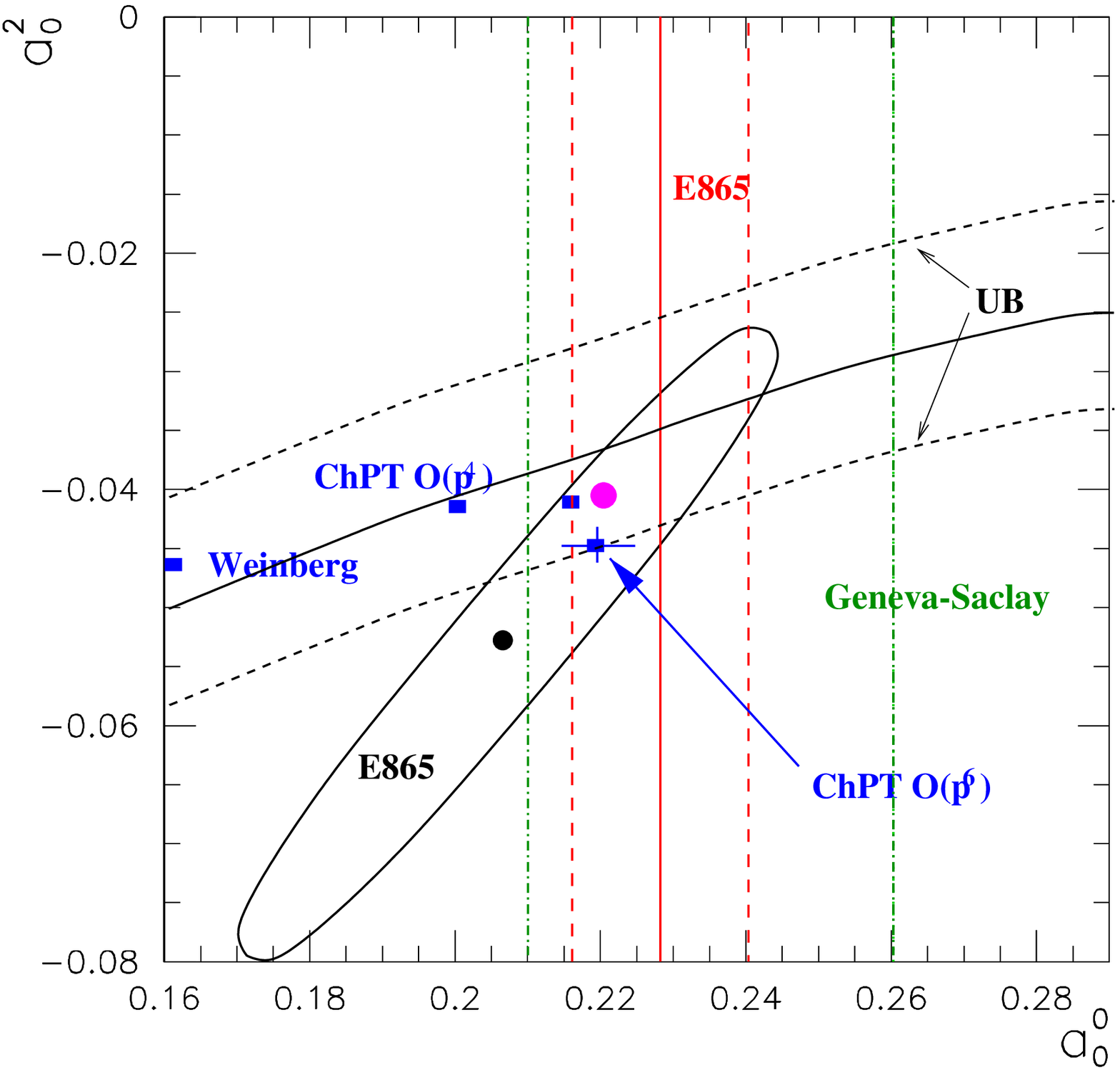}
\caption{\it Left: E865 phase shift results
compared to the Geneva-Saclay data\cite{Rosselet77}. The lower
curve is the result of the fit to our data.
Right: $\pi\pi$ scattering length.
The solid vertical line and the ellipse indicate our results for the
scattering lengths for the two analyses requiring the 
universal curve (UB) as additional input or not. The solid squares
indicate sucessive orders in chiral perturbation theory
calculations, the latest of which~\cite{Colangelo00} is marked
with a cross. The solid circle indicates LQCD-result\cite{Pipith}.
The dot-dashed lines indicate
the center and lower error band of the previous experiment~\cite{Rosselet77}.
\label{fig:pipi4}}
\end{figure}

In a first step the form factors in each bin in
$s_\pi$ are fitted independently. It is further assumed
that the form factors do not depend on
$s_e$ and that the form factor $F$ contributes to $s$-waves only,
i.e. $f_e=g_e=\tilde{f}_p=0$ are fixed at zero. This leaves one parameter
each to describe the form factors $F$, $G$, and $H$ per bin plus
the phase shift difference $\delta\equiv\delta_0^0-\delta_1^1$.
Figures~\ref{fig:pipi3} and \ref{fig:pipi4}
summarize the results for $F$, $G$ and the phase shifts, respectively.
The $s_\pi$ dependence of $F$ and $G$ may be fitted with
$f_s = 5.770\pm0.097$, $f_s^\prime= 0.95\pm 0.55$,
$f_s^{\prime\prime}=-0.52\pm 0.58$ (quadratic fit) or
with $f_s =5.835\pm0.065\ ,\; f_s^\prime= 0.47\pm0.15$
(linear fit), and $g_p = 4.684\pm 0.092$, $g_p^\prime= 0.54\pm0.20$
(linear fit), respectively.

If the individual phase shifts are expressed 
in terms of the scattering length $a^0_0$ (Eq.~\ref{eq:schenk}), 
i.e. the {\em universal curve} is used to
express $a^2_0$ as a function of $a^0_0$, and the form factors
are parametrized as in Eq.~\ref{eq:AB99}, the
results listed in Table~\ref{tab:pipi} are obtained.
\begin{table}[htb]
\caption{\it Results of the fits to the $K_{e4}$ data using
Eq.~\ref{eq:AB99} and Eq.~\ref{eq:UB} ($\chi^2/\mbox{NdF}=1.075$).
For the last column the latter constraint is dropped. The
other parameters and the $\chi^2$ do not change significantly.} 
\vskip 0.1 in
\centering
\begin{tabular}{|lr|lr|}\hline
$f_s$ & $ 5.75\pm0.02 \pm0.08$  & $g_p$ & $ 4.66\pm0.47 \pm0.07$\\  
$f_s^\prime$ & $1.06\pm0.10\pm0.40$ & $g_p^\prime$ & $0.67\pm0.10\pm0.04$\\ 
$f_s^{\prime\prime}$&$-0.59\pm0.11\pm0.40$ &$h_p$& $-2.95\pm0.19\pm0.20$\\
\hline \hline
$a^0_0$&\multicolumn{3}{l|}{$0.228\pm0.012\pm0.003$
$\;\Rightarrow\;$ (UB) $a_0^2\;\;-0.036\pm0.009$} \\ \hline
$a^0_0$ & 
\multicolumn{1}{l|}{$0.203\pm0.033\pm0.004$} 
&$a^2_0$ & $-0.055\pm0.023\pm0.003$ \\\hline
\end{tabular}\label{tab:pipi}  
\end{table}

In conclusion it can be stated, that
the E865 results constitute a major advance in the program
of precision tests of hadronic interaction models
at low energies. They also do nicely confirm
the updated two-loop ChPT predictions\cite{Colangelo00}.


\end{document}